\documentclass[12pt]{article} 

\usepackage{graphicx} 
\PassOptionsToPackage{hyphens}{url}\usepackage{hyperref} 
\pdfoutput=1

\title{Limits to green growth and the dynamics of innovation} 

\author{Salvador Pueyo\textsuperscript{a,b,}\thanks{E-mail: spueyo@riseup.net}\\\textit{\small{\textsuperscript{a}Dept. Evolutionary Biology, Ecology, and Environmental Sciences, }}\\\textit{\small{Universitat de Barcelona, Av. Diagonal 645, 08028 Barcelona,}}\\\textit{\small{Catalonia, Spain}}\\\textit{\small{\textsuperscript{b}Research \& Degrowth, C/ Trafalgar 8 3, 08010 Barcelona, Catalonia, Spain}}} 
\date{} 
\begin{document} 
\maketitle 
\begin{abstract} 
\noindent Central to the official \textit{green growth} discourse is the conjecture that absolute decoupling can be achieved with certain market instruments. This paper evaluates this claim focusing on the role of technology, while changes in GDP composition are treated elsewhere. Some fundamental difficulties for absolute decoupling, referring specifically to thermodynamic costs, are identified through a stylized model based on empirical knowledge on innovation and learning. Normally, monetary costs decrease more slowly than production grows, and this is unlikely to change should monetary costs align with thermodynamic costs, except, potentially, in the transition after the price reform. Furthermore, thermodynamic efficiency must eventually saturate for physical reasons. While this model, as usual, introduces technological innovation just as a source of efficiency, innovation also creates challenges: therefore, attempts to sustain growth by ever-accelerating innovation collide also with the limited reaction capacity of people and institutions. Information technology could disrupt innovation dynamics in the future, permitting quicker gains in eco-efficiency, but only up to saturation and exacerbating the downsides of innovation. These observations suggest that long-term sustainability requires much deeper transformations than the green growth discourse presumes, exposing the need to rethink scales, tempos and institutions, in line with ecological economics and the degrowth literature.
\end{abstract} 


\section{Introduction} 
\label{Introduction}

Will technological progress offset resource exhaustion and neutralize the environmental impacts of growth? The answer given by the simulations in Meadows et al.\ (1972) well-known report on the \textit{Limits to Growth} was \textit{no}. These authors concluded that growth-oriented policies had to be urgently subverted to avoid societal collapse at some point in the 21st century. The political challenge posed by this and many other studies (see, e.g., Daly et al., 1973) was taken by the green movement of the 70s-80s\footnote{See, e.g., the documents in \url{https://grunentodegrowth.wordpress.com/}} and has been more recently retaken with denominations such as, notably, \textit{degrowth} (D'Alisa et al., 2014; Kallis, 2018; Kallis et al., 2018; Hickel, 2019). The mainstream position was quite different; it was defined by two contradictory responses from neoclassical economists: (1) the assertion (emphatic, but without evidence) that technology would offset these negative feedbacks (e.g., Solow, 1973) and (2) the continuing omission of such feedbacks from formal growth models (to which Solow himself was a prominent contributor), in which, when technological progress is switched off, economies just tend to a steady state not followed by decline (Barro and Sala-i-Martin, 2004). Recognizing that the feedbacks do exist, the present paper gives some formal and evidence-based steps to clarify whether or not they are going to be offset. In so doing, it addresses a key point of disagreement between mainstream and ecological economics (e.g., Martinez-Alier, 2013), as is the soundness of long-term economic forecasts and policies that neglect the biophysical dimensions of economies.

The rationale for the offset hypothesis was that resource scarcity would automatically create price signals which would induce the due developments to enable resource efficiency or replacement as needed (Solow, 1973). More recent mainstream literature takes this for granted (even though long-term availability is uncertain and prices are heavily conditioned by other factors; Kallis and Sager, 2017) and focuses its attention on environmental impacts, especially climate change. Under the label of \textit{green growth} it is claimed that such impacts can be tackled with the help of price-based instruments such as taxes and tradable permits\footnote{UNEP (2014) does not count tradable permits as \textit{price-based instruments}, but, e.g., OECD (2011a) does. Here, the second convention is followed.} while economies keep growing; this is currently the dominant discourse of major multilateral institutions (OECD, 2011a; UNEP, 2014). Other instruments are considered, but according to OECD (2011b, p. 12), only when \textit{there are intractable political obstacles to price-based measures}, and in no case questioning growth itself. 

Much of the discussion about green growth revolves around the potential for decoupling GDP from the consumption of energy or materials or from greenhouse gas emissions, in particular whether it is possible to move from \textit{relative decoupling} (with such physical flows growing more slowly than GDP but still growing) to \textit{absolute decoupling} (with physical flows declining while GDP grows). A number of arguments against the plausibility of absolute decoupling have been advanced (e.g., Ward et al., 2016; Kallis, 2017), notably the mere observation that it is not taking place, as is backed by a number of empirical studies which take into account international fluxes (Steinberger et al., 2010, 2013; Bithas and Kalimeris, 2013; Csereklyei and Stern, 2015; Haas et al., 2015; Plank et al., 2018; Cattaneo et al., 2018), with recent data indicating even \textit{recoupling} (Schandl et al., 2018). However, would stronger, generalized application of the due pricing mechanisms permit absolute decoupling? The core results of the present paper concern this point. 

The eco-efficiency of an economy depends not only on the eco-efficiency of each production process but also on GDP composition. The present paper focuses on the former, while the latter will be treated in detail in a separate paper. In principle, absolute decoupling could be reached if economic activity concentrated increasingly in  less resource-intensive sectors (because of changes in consumption, not offshoring, which is highly relevant at the level of specific countries; Steinberger et al., 2010, 2012; Plank et al., 2018; Cattaneo et al., 2018; Hardt et al., 2018). However, the viability of this process is not entirely clear (sec.\ \ref{decoupling}) and, if it were driven by prices (either because of green growth's market instruments or because of resource depletion), it would compromise the access of lower-income strata to the resources needed to satisfy their basic needs. This would contradict the purported goal to reduce poverty, which is a fundamental argument to justify the \textit{growth} component of the \textit{green growth} binome (e.g., Smulders et al., 2014). The green growth toolkit is ill-equipped to deal with this challenge, since it takes for granted mainstream economic policies except in the quest to increase the use of instruments favoring eco-efficiency, which is probably the reason why it also takes growth as the default tool to fight poverty. In this aspect it differs sharply from degrowth proposals, which emphasize structural changes in the management and allocation of resources. \textit{Degrowth} does not mean negative growth but an exit from the logic of GDP growth to prioritize long-term social goals within environmental limits, whichever the implications for GDP, but assuming that this process will probably involve some period of negative growth, at least in high-income countries (D'Alisa et al., 2014; Kallis, 2018; Kallis et al., 2018). 

Section \ref{foundations} develops a stylized model of the dynamics of thermodynamic efficiency based on the available knowledge on the dynamics of learning and innovation in general. This builds on Ward et al.'s (2016) and Magee and Deveza's (2017, 2018) previous work along related lines (compared in sec.\ \ref{decoupling}). Section \ref{results} uses the model to tentatively identify the conditions for sustained decreases in exergy consumption (or, proportionally, entropy production) to take place, considering positive, zero and negative growth. Within the Discussion (sec.\ \ref{discussion}), sec.\ \ref{decoupling} interprets the results, and sec.\ \ref{energy} explores their practical relevance for the debate on limits. Section \ref{acceleration} notes that increasing efficiency cannot be dissociated from other dimensions of technological innovation, such as the creation of new environmental and social challenges, as well as mutations in the very patterns of innovation, and discusses the implications for the feasibility of green growth. Section \ref{conclusions} concludes. 

\section{Model} 
\label{foundations} 

The steps to find the maximum of a function, such as some measure of efficiency, cannot always be defined a priori. This is the case when a problem combines features such as computational intractability (Arora \& Barak, 2009), partial information and open-ended possibilities. The many steps that both biological and cultural evolution needed to reach some given solutions, and the relatively \textit{wandering} trajectories that led to such solutions, attest to the practical centrality of this kind of problems. In such cases it is not possible to predict when some specific advances will take place, but, when referring to more generic forecasts, empirical observations of innovation and learning dynamics permit some educated guesses. It is on such guesses that the present paper stands. 

Traditionally, firms do not seek reductions in natural resource or emission intensity \textit{per se}, but reductions in monetary cost, of which the cost of natural resources is just one component. However, knowledge on innovation to reduce monetary costs can help to characterize the dynamics of innovation more generally.  Unfortunately, even in this well-documented context, large uncertainties remain. A number of models exist (Anzanello and Fogliatto, 2011) for what is variously termed \textit{experience curves}, \textit{learning curves} (K{\"o}ler et al., 2006) or \textit{progress curves} (Dutton and Thomas, 1984). Such curves capture a combination of novel knowledge (either technological or of other kinds, either resulting from experience or from R{\&}D investments), implementation of such knowledge and gains in workers' skills. Two especially popular models are Wright's and Moore's. Besides their popularity, Nagy et al.'s (2013) analysis of 62 time series for different products gave statistical evidence of the superiority of these two models over several alternatives, with Wright's performing marginally better than Moore's. 
 
According to Wright's rule (better known as \textit{Wright's law}), costs $c$ decrease as a function of cumulative production $Q$ following approximately a power law: 
\begin{equation} 
\label{Wright} 
c_t = a Q_t^{-\lambda}, 
\end{equation} 
where $t$ is time and $a$, $\lambda$ are assumed constant and positive. The rule was formulated by Wright (1936) for airplane production, referring specifically to labor costs and also to (monetized) material costs, which he found to be much less responsive to the accumulated experience (smaller $\lambda$ than for labor costs). Thereafter, it has been documented in many instances, mostly referring to full monetary costs, both at the firm (e.g., Dutton and Thomas, 1984) and industry level (e.g., Weiss et al., 2010; Nagy et al., 2013). In sustainability science, Wright's rule has been widely used to estimate future monetary costs of different sources of energy, adding in some cases the assumption of a floor cost that cannot be trespassed (K{\"o}ler et al., 2006). With variables other than monetary cost, the same equation has been widely applied in psychology to fit data of individual learning (e.g., Newell and Rosenbloom, 1981). 

According to Moore's rule (better known as \textit{Moore's law}) when applied to costs, these would decrease exponentially over time $t$: 
\begin{equation} 
\label{Moore} 
c_t = a e^{-ht}, 
\end{equation} 
where $a$, $h$ are also constants expected to be positive. Wright's and Moore's equations can be interpreted as different kinds of innovation, endogenous in the first case and exogenous in the second (Nordhaus, 2014), but this distinction is often difficult to grasp from empirical data because Eqs.\ \ref{Wright} and \ref{Moore} are equivalent when production increases exponentially (Sahal, 1979). Since both forms of innovation are thought to operate, authors such as Nordhaus (2014) combine both rules: 
\begin{equation} 
\label{Nordhaus} 
c_t = a Q_t^{-\lambda} e^{-ht}. 
\end{equation} 
Nordhaus also assumes that there is bidirectional causality, from cumulative production to cost and from declining cost to production growth (i.e., rebound effect; Schneider, 2008), and shows that Eq.\ \ref{Nordhaus} is consistent with this assumption. In this paper, it will be assumed that Eq.\ \ref{Nordhaus} is generally valid, with $\lambda < 1$. This assumption about $\lambda$, which implies diminishing returns, is justified in the Appendix, and is more relevant for the results than the precise shape of Eq.\ \ref{Nordhaus}.

Consider a scenario in which monetary prices are more aligned with resource or environmental costs, or, in a theoretical extreme, the former are entirely determined by the latter. Would the above relation still apply, and serve to forecast changes in eco-efficiency? The answer will depend on the weight given to different dimensions of eco-efficiency. For example, CFCs were phased-out by the Montreal Protocol without major disruptions, and, even thought they were largely replaced by other halocarbons (Tollefson, 2009), it was quickly shown that, vested interests aside, nothing prevented the entire elimination of halocarbons at least in refrigerators (Conrad, 1995). Market signals appear to be less effective than regulation in this case (Tollefson, 2009), but both regulation and strong enough market signals should have been able to terminate halocarbon use in this sector abruptly, rather than inducing an endless sequence of limited reductions in the \textit{halocarbon cost} of refrigeration in agreement to some rule such as Eq.\ \ref{Nordhaus}. 

In the scenarios explored in this paper, monetary cost is aligned, specifically, with thermodynamic cost, i.e., with the exergy spent to carry out a given economic activity, directly or indirectly, exosomatically or endosomatically (the spent exergy or \textit{exergy consumption} is closely related to what is normally called \textit{energy consumption} but is a more precise measure, and is proportional to entropy production (Dincer and Rosen, 2013); its relevance for sustainability is discussed in sec.\ \ref{energy}). It is argued below that, in the considered scenarios, reductions in thermodynamic cost should differ sharply from the halocarbon instance and have features analogous to reductions in monetary cost. Already with current price criteria, Wright's equation (Eq.\ \ref{Wright}) with $\lambda < 1$ has been fitted to a few production processes in terms of energy instead of monetary costs (Ram{\'i}rez and Worrell, 2006) and to many energy saving technologies in terms of monetary costs (Weiss et al., 2010). These studies are encouraging, but the main arguments for this extension are more fundamental. 

While the theoretical possibility exists to develop technologies to dispense with specific resources or emissions, the laws of thermodynamics imply that thermodynamic costs are and will always be nonzero for all economic actions (Pueyo, 2003, p. 307)\footnote{This is true provided that, when some given expenditure of exergy (either exosomatic or endosomatic) serves multiple purposes, none of them is assigned 0\% of the shared thermodynamic cost. Strictly speaking, the results in this paper apply to accounting criteria that do not violate this principle}. Not even the need for labor is so fundamental. While conventional firms currently seek to reduce overall costs by addressing the monetary cost of each of the elements that enters the production process (by refining the techniques used in each of the production steps that take place within the specific industry and also in their choice of inputs, which, on its turn, sends signals up the supply chain or network), in the scenarios here considered they would change criteria and attempt to reduce the thermodynamic costs of, at least, this same set of elements. Therefore, the problem would be similar, with at least the same elements and the same structure of connections among them, not just within each firm but throughout the supply network. What would change would be the weight given to each of these elements, which would remain positive in all cases, and the possible addition of even more elements which currently have no monetary cost. Generally, the complexity should be expected to be the same or greater. Since Eq.\ \ref{Nordhaus} or some similar rule seems to apply to most if not all production processes despite their great heterogeneity, and the elements entering such processes do so with extremely heterogeneous weights, the assumed changes are unlikely to alter this rule, except in that, in the thermodynamic case, it is certain that there is some nonzero floor cost (a connection between thermodynamics and bounds to learning curves was already suggested by Baumg{\"a}rtner, 2002).

Let us thus adapt Eq.\ \ref{Nordhaus} for the application intended in this paper. A simple way to introduce a floor cost $c_f$ (now referring to thermodynamic cost) is as an upward translation of $c$: 
\begin{equation} 
\label{withfloor0} 
c_t = c_f + a Q_t^{-\lambda} e^{-h t}. 
\end{equation} 
This is indistinguishable from Eq.\ \ref{Nordhaus} when $c_f << c_t$. If exogenous innovation is excluded ($h=0$), it has the same form as a type of equation widely used in statistical physics to describe the relation between different variables when approaching a critical threshold (Garrod, 1995). However, this model is still insufficient because, like Eqs.\ \ref{Wright}-\ref{Nordhaus}, Eq.\ \ref{withfloor0} presupposes some degree of constancy in the parameters, while this paper deals with policy changes that increase the weight of thermodynamic costs in monetary costs. Let us, thus, define $t=0$ as the point in time at which the policy package is implemented. At $t=0$, there will be more opportunities for improvement than there would be if the policy package had always been in place, case in which the same wealth of opportunities would have been found at some $t<0$, when $Q$ was lower than $Q_0$. Therefore, the impact of the policy change will be tentatively introduced by replacing $Q_0$ by the effective value $Q_0^e$ ($Q_0^e \leq Q_0$) that makes Eq.\ \ref{withfloor0} consistent with the thermodynamic cost $c_0$, i.e., 
\begin{equation} 
\label{withfloor} 
c_t = c_f + a (Q_0^e + [Q_t -Q_0])^{-\lambda} e^{-h t}. 
\end{equation} 

Some aspects of this model need clarification. First, its output must be interpreted as a lower bound to exergy consumption, because it does not include the added thermodynamic cost of resource extraction as scarcity increases (Bardi, 2013), as well as possible costs of mitigation of and adaptation to environmental degradation. Second, Eq.\ \ref{withfloor} is a smooth function, while, in the psychological arena, Donner and Hardy (2015) found that individual learning curves were punctuated by abrupt shifts, with only averages being well fitted by Eq.\ \ref{Wright}. Also the collective accumulation of technical knowledge is punctuated by key inventions, so Eq.\ \ref{withfloor} should be interpreted as a model for the average curve, but this is precislely the most relevant to evaluate the viability and convenience of green growth. The mentioned results by Nagy et al. (2013) were obtained by putting quite different technologies together when those performed the same function. Third, the specific mathematical choices to adapt Eq.\ \ref{Nordhaus} to this context, which result in Eq.\ \ref{withfloor}, are tentative and in need of empirical evaluation. Fourth, innovation rates could change nonlinearly in the future, which is discussed in sec.\ \ref{acceleration}.

\section{Analysis} 
\label{results} 

GDP growing exponentially at rate $r$ while keeping its composition constant is equivalent to each sector's production $y$ growing at rate $r$; let us thus focus on a single sector. This section gives some relations between production growth and exergy consumption that follow mathematically from Eq.\ \ref{withfloor} with $\lambda < 1$. Let the policy package aligning monetary with thermodynamic costs be applied at time $t=0$. Let, thereafter, production $y$ grow (or decline) at rate $r$, with $r \neq 0$ (the case $r=0$ is given special treatment below). Then, production at time $t$ will be 
\begin{equation} 
\label{exponencial} 
y_t = y_0 e^{rt}. 
\end{equation} 
The corresponding cumulative production $Q$ is: 
\begin{equation} 
\label{Q} 
Q_t = Q_0 + r^{-1} y_0 (e^{rt} -1), 
\end{equation} 
which, combined with Eq.\ \ref{withfloor}, gives 
\begin{equation} 
\label{cost1} 
c_t = c_f + a (Q_0^e + r^{-1} y_0 (e^{rt} - 1))^{-\lambda} e^{-h t}. 
\end{equation} 
Let us define the policy impact coefficient 
\begin{equation} 
\label{prepolicy1} 
p=\frac{Q_0}{Q_0^e}
\end{equation}
and
\begin{equation}
\label{rb}
r_b=\frac{y_0}{Q_0}
\end{equation}
 ($r_b > 0$), with $r_b$ being an integrated measure of the sequence of growth rates before $t=0$. The rationale for Eq.\ \ref{rb} is that, if before $t = 0$ production had grown at a constant rate for an asymptotically long period, this rate would be $r_b$. Combining Eqs.\ \ref{prepolicy1}-\ref{rb},
\begin{equation} 
\label{policy1}
Q_0^e = p^{-1} r_b^{-1} y_0.
\end{equation}
From Eqs.\ \ref{cost1} and \ref{policy1}, the thermodynamic cost is 
\begin{equation} 
\label{cost2} 
c_t = c_f + a y_0^{-\lambda} ([ p^{-1} r_b^{-1}-r^{-1}] + r^{-1}e^{rt}) ^{-\lambda} e^{-h t}. 
\end{equation} 
Exergy consumption (proportional to entropy production) is 
\begin{equation} 
\label{basica} 
E_t = c_t y_t. 
\end{equation} 
From Eqs.\ \ref{exponencial}, \ref{cost2} and \ref{basica},
\begin{equation}
\label{exergia_abs}
E_t = c_f y_0 e^{rt} + a y_0^{1-\lambda} ([ p^{-1} r_b^{-1}-r^{-1}] + r^{-1}e^{rt}) ^{-\lambda} e^{(r-h) t}.
\end{equation}
These equations can be expressed more simply as a ratio to the initial values. Let us define $\theta$ as the physical margin to reduce thermodynamic costs,
\begin{equation}
\label{teta}
\theta_t = \frac{c_f}{c_t}, 
\end{equation}
so $c_0 = (1-\theta_0 )^{-1} (c_0 - c_f)$. From Eqs.\ \ref{cost2} and \ref{teta},
\begin{equation} 
\label{ratiocost} 
\frac{c_t}{c_0} = \theta_0 + ( 1-\theta_0 ) \left( 1 + \left[ p \frac{r_b}{r} \right] (e^{rt}-1) \right) ^{-\lambda} e^{-h t}. 
\end{equation} 
Similarly, defining $\epsilon_t = \frac{E_t}{E_0}$, Eq.\ \ref{exergia_abs} becomes
\begin{equation} 
\label{exergia} 
\epsilon_t = \theta_0 e^{rt} + ( 1-\theta_0 ) \left( 1 + \left[ p \frac{r_b}{r} \right] (e^{rt}-1) \right)^{-\lambda} e^{(r-h) t}. 
\end{equation} 
Exergy consumption will decrease\footnote{Exergy consumption will decrease, $\frac{dE_t}{dt} < 0$, if and only if $\frac{1}{\epsilon_t - \epsilon_f} \frac{d \epsilon_t}{dt} < 0$, where $\epsilon_f = \theta_0 e^{rt}$, which gives Eq.\ \ref{condicio1} from Eq.\ \ref{exergia}.} if and only if 
\begin{equation} 
\label{condicio1}
r-h-\lambda r \left( \left[ p^{-1} \frac{r}{r_b}-1 \right] e^{-rt}+1 \right) ^{-1} + r ( \theta_t^{-1}-1 )^{-1} < 0. 
\end{equation} 
A consequence is that decreasing production results in decreasing exergy consumption in all cases: For $r<0$, the first term in Eq.\ \ref{condicio1} is negative and the others are $\leq 0$, giving a negative sum.

Let us move to the case of growing production ($r>0$), which is more complex because of the sum of positive and negative terms in Eq.\ \ref{condicio1}. Initially ($t=0$), exergy consumption will decrease if and only if 
\begin{equation} 
\label{condicio2} 
r-h-\lambda r_b p + r ( \theta_0^{-1}-1 )^{-1} < 0. 
\end{equation} 
This equation indicates that, even for large growth rates $r$, a large enough $p$ would initially result in absolute decoupling, if $\theta_0 << 1$. However, this effect would be short-lived. Only if $r$ approaches zero (or is negative) does $p$ have a long-lasting impact. Otherwise, the policy shock vanishes exponentially, giving
\begin{equation}
\label{assimptotica_abs}
E_t \approx c_f y_0 e^{rt} + a y_0^{1-\lambda} r^{\lambda} e^{[(1-\lambda)r-h] t},
\end{equation}
or, in relative terms,
\begin{equation} 
\label{assimptotica} 
\epsilon_t \approx \theta_0 e^{rt} + ( 1-\theta_0 )\left[ p \frac{r_b}{r} \right]^{-\lambda} e^{[(1-\lambda) r-h] t}
\end{equation} 
(where $p$ and $r_b$ remain relevant only because of the division by $E_0$). At this stage, exergy consumption decreases if and only if 
\begin{equation}
\label{condicio3} 
r(1-\lambda) -h + r ( \theta_t^{-1}-1 )^{-1} < 0. 
\end{equation} 

A consequence of Eq.\ \ref{condicio3} is that endogenous innovation alone is unable to sustain absolute decoupling, since this inequality is not satisfied for $h=0$ (given $r>0$). Would ordinary levels of exogenous innovation suffice to achieve absolute decoupling? From U.S. data, Nordhaus (2014) estimated a typical $h \approx 0.01$ y$^{-1}$ (for monetary costs). As a representative endogenous innovation parameter, let us take $\lambda \approx 1/3$ (corresponding to the \textit{80\% progress curve}, which became standard in the literature since first fitted by Wright (1936) to labor costs in airplane production, and is far more optimistic than the value $\lambda = 0.0732$, or 95\% curve, which Wright estimated for raw material costs in the same context). Then, the \textit{breakeven} rate of growth in which exergy consumption would neither increase nor decrease would lie at some point between $r=0$ and $r \approx 0.015$ y$^{-1}$ (0 to 1.5 \% per year; i.e., between -1 and 0.5\% y$^{-1}$ on a per capita basis should population grow at 1\% y$^{-1}$), depending on $\theta$. This suggests that, apart from a short period after the policy shock, absolute decoupling would either be unfeasible or imply rates of growth that are generally considered small, and also quite small rates of exergy reduction. 

The condition for production growth closer to standard expectations without increasing $\epsilon$, or, alternatively, more substantial exergy reductions without some production growth, would be a greater $h$, which, in principle, could be attained by a boost in public research funding. From Eq.\ \ref{condicio1}, the $h$ needed for constant $\epsilon$ given a constant $r>0$ (implying $\theta_t = \theta_0 e^{rt}$) would be
\begin{equation}
\label{hdinamic}
h_t^{d} = r -\lambda r \left( \left[ p^{-1} \frac{r}{r_b}-1 \right] e^{-rt}+1 \right) ^{-1} + r ( \theta_0^{-1} e^{-rt}-1 )^{-1}
\end{equation}
or, after the policy shock vanishes,
\[
h_t^{d} = r (1-\lambda) + r ( \theta_0^{-1} e^{-rt}-1 )^{-1}.
\]
However, the required boost in public funding would meet increasing obstacles (discussed in sec.\ \ref{decoupling}) and would become impossible before (possibly well before)
\begin{equation}
\label{criticaltime}
t_{\infty}=-\frac{\ln(\theta_0)}{r},
\end{equation}
when $h \rightarrow \infty$ would be required. At that point, also relative decoupling would be impossible.

The case of zero growth ($r=0$) needs special treatment because Eq.\ \ref{Q} does not apply. In this case, for all $t>0$, $y_t=y_0$ and, therefore $Q_t = Q_0 + y_0 t$. Introducing this expression in Eq.\ \ref{withfloor} and combining with Eqs.\ \ref{basica} and \ref{policy1}, 
\[
c_t = c_f + ay_0^{-\lambda} ([pr_b]^{-1} +t)^{-\lambda} e^{-ht} 
\]
and 
\[ 
E_t = c_f y_0 + a y_0^{1-\lambda} ([p r_b]^{-1} + t)^{-\lambda} e^{-ht}, 
\] 
or, equivalently, 
\begin{equation} 
\label{estacionaria} 
\epsilon_t = \theta_0 + (1-\theta_0 ) (1 + [p r_b] t)^{-\lambda} e^{-ht}. 
\end{equation} 
These are decreasing functions (unless $\lambda = 0$ and $h=0$, in which case they are constant), with $\epsilon$ tending asymptotically to $\theta_0$. 

\begin{figure}[t] 
\makebox[\textwidth][c]{\includegraphics[scale=0.7]{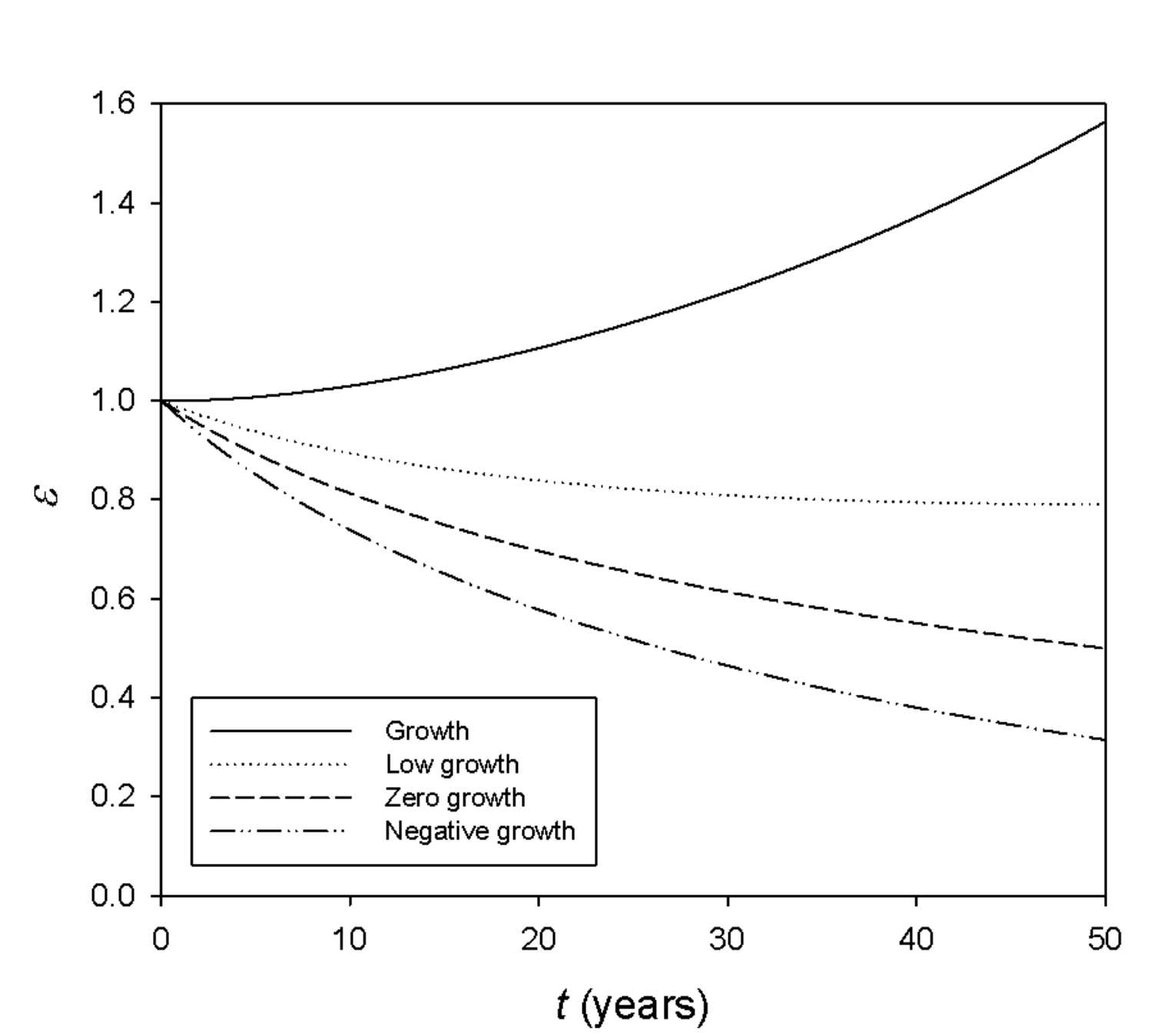}}
\caption{Example of trajectories of exergy consumption in a one-sector economy for different growth trajectories, according to Eqs.\ \ref{exergia} and \ref{estacionaria}, with strong environmental policies being implemented at time $t=0$. The x-axis is time in years, and the y-axis is the ratio $\epsilon$ between current and initial exergy consumption. Parameter values (sec.\ \ref{results}): $\lambda = 1/3$, $h=1\%$ y$^{-1}$, $\theta_0 = 0.2$, $p = 2.6$, $r_b = 2.5\% y^{-1}$. Scenarios: \textbf{Growth} (growth rate $r=2.5\%$ y$^{-1}$), \textbf{Low growth} ($r=1\%$ y$^{-1}$), \textbf{Zero growth} ($r=0$), \textbf{Negative growth} ($r=-1\%$ y$^{-1}$; these are absolute, not per capita growth rates).  These results do not include (1) the increasing thermodynamic costs of extracting resources as they are depleted, as well as costs due to environmental degradation, (2) the rebound effect and (3) 2nd-order innovation; these factors are discussed separately.} 
\label{fig1} 
\end{figure}

Figure \ref{fig1} gives, as an example, a set of exergy consumption trajectories differing in $r$ but sharing the  other parameter values, calculated from Eqs.\ \ref{exergia} and \ref{estacionaria}. As above, $\lambda = 1/3$ and $h = 0.01$ y$^{-1}$. The parameter $\theta$ is difficult to estimate and should differ substantially for different cases; for this example, $\theta_0 = 0.2$ was chosen, meaning that there is the potential to reduce the thermodynamic cost by 80\%. The growth rate assumed to precede the new policy was $r_b = 0.025$ y$^{-1}$. Choosing a plausible $p$ is not straightforward: this parameter defines the magnitude of the \textit{low hanging fruits} that permit a quick increase in efficiency right after the onset of the new market signals. For $r>0$, this is due to thermodynamic costs moving away from the trajectory defined by Eq.\ \ref{withfloor0} with $Q_0$ replaced by $Q_0^e$ and converging to the trajectory defined by the same equation with $Q_0$ replaced by $r^{-1} y_0$. Even thought the conceptual connection is vague, for these examples $p$ was decided by taking the case of unchanging growth rate, $r=r_b$, and assigning a value to $\gamma = \frac{c_0 (Q_0^e ) - c_0 ( p Q_0^e )}{c_0 (Q_0^e )}$ (with $c_0$ calculated from Eq.\ \ref{withfloor0} and using Eq.\ \ref{policy1}) in agreement to the worldwide \textit{savings from adoption of best practice commercial technologies in manufacturing industries} estimated by IEA (2007, p. 35), of 18 - 26 \%. From Eq.\ \ref{withfloor} and the definition of $\gamma$, 
\[ 
p = \left( 1 - \frac{\gamma}{1 - \theta_0} \right) ^{-\frac{1}{\lambda}}, 
\] 
where $c_0$ is the real initial cost $c_0 = c_ 0 (Q_0^e)$ (this expression reduces to $p = (1 - \gamma )^{-\frac{1}{\lambda}}$ for $\theta_0 \rightarrow 0 $). The mid value of IEA's range, $\gamma = 0.22$, corresponds to $p \approx 2.6$, which was the assigned value. The scenarios are: \textit{growth} (with a relatively modest $r = 0.025$ y$^{-1}$), \textit{low growth} ($r = 0.01$ y$^{-1}$; i.e., $r=h$), \textit{zero growth} ($r = 0$), and \textit{negative growth} ($r = -0.01$ y$^{-1}$). In the \textit{growth} scenario, exergy consumption $\epsilon$ starts almost flat, as the leap to 11\% lower $c$ in the first five years almost offsets the 13\% increase of $y$ in this period, where $\epsilon$ increases just by 0.8\%. The 22\% reduction in $c$ corresponding to $\gamma$ takes 11.45 years, at which point $\epsilon$ has increased by 3.8\%. After 25 years, $\epsilon$ has increased by 16\%, and by 56\% after 50 years, while $c$ has been reduced just by 55\%, still far from the 80\% allowed by physics in this example, i.e., ($1-\theta_0$). In the \textit{low growth} scenario, there is absolute decoupling, with $\epsilon$ first decreasing but then stagnating at about 20\% below the initial value. At the end of the series the rate of change in $\epsilon$ is marginally positive, anticipating a return to noticeably positive rates as further decreases in $c$ become increasingly difficult. Both \textit{zero growth} and \textit{negative growth} permit sustained decreases in $\epsilon$; at the end of the series, these amount to 50\% in the first case and 69\% in the second.

\begin{figure}[t] 
\makebox[\textwidth][c]{\includegraphics[scale=0.6]{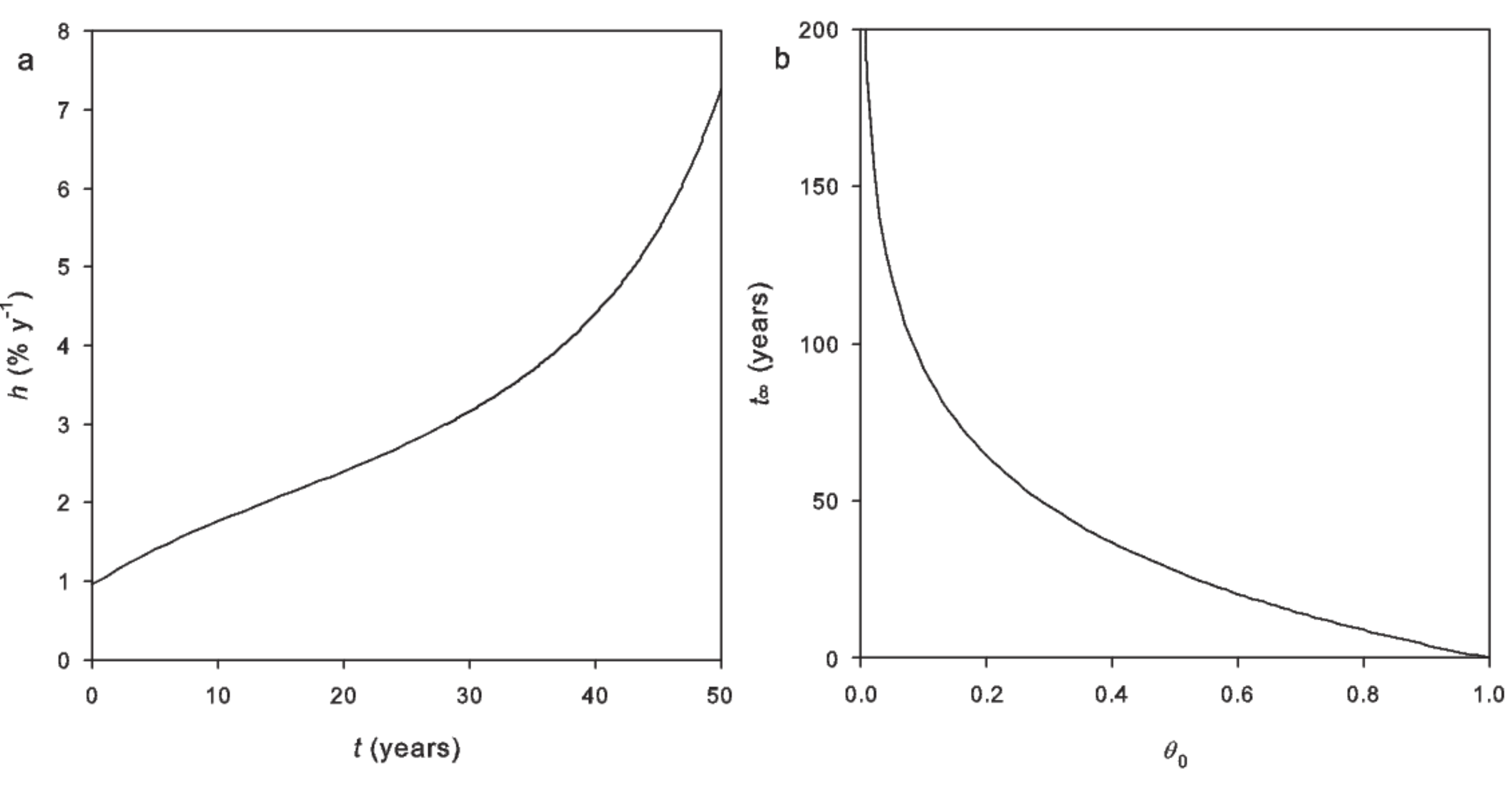}}
\caption{Limits to decoupling by increasing exogenous innovation (e.g., with public research funding). a: Trajectory of $h^d$ (\% y$^{-1}$) that, if it were feasible, would keep exergy consumption constant (Eq.\ \ref{hdinamic}), for the parameter values of the \textit{growth} scenario (except $h$), in which production grows at 2.5\% y$^{-1}$ and physics permits a decrease in thermodynamic costs of 80\% ($\theta_0 = 0.2$). At $t_\infty = 64.4$ years, $h_t^{d} \rightarrow \infty$ would be needed. Note that $h^d$ is itself a rate of exponential increase, of what can be roughly interpreted as the amount of exogenous innovations per unit time. b: Time $t_\infty$ (years) at which $h_t^{d} \rightarrow \infty$ would be needed to keep exergy consumption constant, also for growth rate $r=2.5$\% y$^{-1}$, but for different values of $\theta_0$ (the other parameters have no effect). The plot does not cover all the asymptote $\theta_0 \rightarrow 0$, $t_\infty \rightarrow \infty$.}
\label{fig2} 
\end{figure}

Figure \ref{fig2} gives the trajectory of $h^d$, defined as the minimum $h$ to keep absolute decoupling (Eq.\ \ref{hdinamic}), for the parameter values in the \textit{growth} scenario above (except $h$). This value (which is itself a rate of exponential increase) would need to double in the first 14 years. After 23 years (with the policy shock having vanished), $h_t^{d}$ starts to grow superexponentially, reaching $h_t^{d} \rightarrow \infty$ at $t_\infty = 64.4$ years (Eq.\ \ref{criticaltime}). Figure \ref{fig2}b shows how $t_\infty$ changes as a function of $\theta_0$ (Eq.\ \ref{criticaltime}), for the same $r$ (this result is independent of $\lambda$, $p$ and $r_b$). A trajectory like that in Fig.\ \ref{fig2}a should become unfeasible at some point before (possibly well before) $t_\infty$.

\section{Discussion} 
\label{discussion} 

\subsection{Is absolute decoupling feasible?} 
\label{decoupling} 

The aim of the stylized model in this paper is to synthesize the available empirical knowledge on the dynamics of innovation and learning and extrapolate it to a hypothetical green growth scenario, where a one-sector economy grows exponentially in a context that is capitalist but with market signals that align prices with thermodynamic costs, as well as nongrowth scenarios. Generic references to thermodynamics were central to foundational discussions on the limits to growth (Soddy, 1922; Georgescu-Roegen, 1971; see also Baumg{\"a}rtner, 2002). The present paper uses a specific model (Eq.\ \ref{withfloor}) to study the potential to decouple production growth from growth in exergy consumption or entropy production, which are closely related to \textit{energy production} but have more precise physical meanings.

Normally, monetary costs decrease more slowly than production grows, because of mechanisms that are likely to operate similarly if monetary costs align with thermodynamic costs (sec.\ \ref{foundations}). As a first approximation, innovation dynamics obey Wright's rule (Eq.\ \ref{Wright} with $\lambda<1$), as apparent from a number of empirical studies (sec.\ \ref{Introduction}). Even though this rule contemplates no physical limit to efficiency, it is incompatible with absolute decoupling, since it predicts a positive growth rate of exergy consumption, ($r(1-\lambda)>0$)\footnote{The growth in relative exergy consumption consistent with Wright's rule is found from Eq.\ \ref{assimptotica} by removing physical limits to efficiency ($\theta_0=0$), exogenous innovation ($h=0$) and transient effects ($p=1$, $r_b = r$), which results in $\epsilon_t = e^{r(1-\lambda)t}$ for $r>0$.}, for production growth $r>0$. However, some departures from this rule can be considered. The model in this paper (Eq.\ \ref{withfloor}) adds two factors that make absolute decoupling more likely, namely, transient effects and exogenous innovation, and one factor that constrains it: the unavoidable physical limits to efficiency. Other important factors that make decoupling more unlikely were left aside, such as the rising thermodynamic costs of extracting resources as these are depleted (Bardi, 2013) and of mitigating or adapting to environmental degradation, and the rebound effect (Schneider, 2008), so, probably, green growth is even less plausible than the model suggests. A phenomenon of potential future importance, 2nd-order innovation, is discussed in sec.\ \ref{acceleration}. The model's forecast for \textit{green} growth is, roughly speaking, a triphasic trajectory with these stages:\\ 

\noindent I- Initial response to the new market signals, resulting in an especially high degree of decoupling, potentially absolute. 

\noindent II- After the initial pulse dissipates, in no case can absolute decoupling be sustained by endogenous innovation alone, and ordinary levels of exogenous innovation are unlikely to suffice. 

\noindent III- Asymptotically, efficiency approaches its physical limit, there is no absolute decoupling and even relative decoupling tends to vanish.\\ 

The transient character of the initial response to new policies (phase I) is consistent with Hardt et al.'s (2018) observation of a substantial deceleration in the gains of thermodynamic efficiency in the UK in recent years. Examples like this could help to improve the quantification of the policy impact coefficient $p$ (Eq.\ \ref{policy1}) for different policies, which is important because the practical implications of this work will differ depending on the potential depth and breath of phase I. The low share of energy in production costs is sometimes thought to imply that firms could reduce energy consumption substantially with little consequence (major $p$ could be easily reached), but strong arguments have been given against this interpretation (Ayres et al., 2013; K{\"u}mmel and Lindenberger, 2014). 

In phase II, after the transient effects have dissipated, only a high enough value of the exogenous innovation parameter $h$ (as compared to $r$) would permit absolute decoupling. Standard values of $h$ are unlikely to suffice, as apparent from the fact that, even in empirical studies that treat all innovation as endogenous (with the consequent overestimation of $\lambda$; Nordhaus, 2014), the results are mostly consistent with $\lambda <1$ (Appendix), and also from the numerical example in sec.\ \ref{results} with plausible parameter values. Therefore, absolute decoupling would probably need a substantial increase in $h$, and would be temporary, because $h$, which is itself a rate of increase, would have to increase superexponentially beyond some point (Eq.\ \ref{hdinamic}, Fig. 2). A means exists to boost $h$, as is public research funding, but any attempt to rise $h$ would meet a number of successive obstacles. First, the general difficulties to provide a public good, possibly magnified by the neoliberal ideas and interests that the green growth project represents (see Dale et al., 2016). Then, economic incompatibility with growth as other investment would be crowded out, inviability for lack of means, and, finally, futility when efficiency saturates for physical reasons ($\theta \rightarrow 1$ in phase III). Another fundamental problem of increasing $h$ is treated in sec.\ \ref{acceleration}.

These tentative conclusions refer to a one-sector economy, and would also apply to complete economies if GDP composition did not change. This is not so clear once composition changes are taken into account, but, continuing the comparison with non-environmental costs, the empirical fact is that economy-wide total factor productivity does increase more slowly than GDP (Barro and Sala-i-Martin, 2004, p.\ 439-440; assuming that the quantitative limitations of Solow's measure of \textit{total factor productivity} do not alter the qualitative conclusion). However, this could result from the rebound effect, which also poses a serious problem to green growth, but not necessarily a definitive one if this were the only issue (see below). Yet, even if it is technically possible to achieve absolute decoupling by inducing changes in GDP composition, this strategy could have serious drawbacks (sec.\ \ref{Introduction}), which are analyzed in detail in a separate paper. 

Scenarios without growth give the best outcomes in sec.\ \ref{results}. These assume that Eq.\ \ref{withfloor} applies also in the absence of production growth, due to the implicit premise that the stabilization or contraction of production takes place in a controlled way, with a more selective allocation of resources with social criteria (rather than being forced by a recession), as is sought by degrowth proposals (D'Alisa et al., 2014; Kallis, 2018; Kallis et al., 2018; Hickel, 2019; analyzed in Pueyo, 2014). These results should be taken with special caution because the deep institutional changes implied by degrowth could modify innovation patterns in one or another way. Furthermore, the improvement could be overestimated if there are benefits of scale playing a direct role in efficiency. On the other hand, the extra exergy costs due to resource depletion and environmental degradation should be lower in nongrowing economies. In the model as it stands, the initial pulse does not dissipate and is followed by continuing increases of efficiency (albeit at a lower rate than in a growing economy), so there is a sustained reduction in exergy consumption.

The model in this paper (Eq.\ \ref{withfloor}) is a first approximation, based on empirical evidence but indirectly. It would benefit from econometric analyses designed to test, refine and, especially, parameterize it. One specific feature of the model, which is the condition $c_f > 0$, needs no econometric analysis because it is imposed by the 2nd Law of thermodynamics (Pueyo, 2003, p. 307). 

Ward et al.\ (2016) and Magee and Deveza (2017, 2018) also obtained results unfavorable to absolute decoupling using learning curves. The novelties in this paper are: (1) While those studies are based on Eq.\ \ref{Moore}, which applies most properly to exogenus innovation, here both exogenous and endogenous innovation are considered (Eq.\  \ref{withfloor}). (2) In line with Ward et al.\ (2016) but differently from Magee and Deveza's (2017, 2018), floor costs are introduced . (3) While those previous studies refer to \textit{business as usual} scenarios, this paper deals with green growth and nongrowth scenarios. (4) In Magee and Deveza (2017, 2018) growth is endogenous, and decoupling fails because of the rebound effect, while this paper addresses the thermodynamic cost of some given, exogenous growth trajectories. The rebound effect has been signaled as a key mechanism making the current economy unsustainable and as a serious obstacle to green growth (Schneider, 2008). It is not the object of the above analysis because it has already been quite studied and because green growth proposals contemplate instruments (OECD, 2011b) which, if applied thoroughly enough, should serve to cap total resource use. The analysis in sec.\ \ref{results} serves to explore if growth would be compatible with such a thorough application of such instruments, or with advanced levels of resource depletion, and gives serious reasons for skepticism. This analysis must be complemented by considering possible future changes in innovation patterns, which are discussed in sec.\ \ref{acceleration}.

\subsection{Relevance of exergy consumption} 
\label{energy} 

The measure of thermodynamic efficiency is one among many possible measures of eco-efficiency, chosen because of the especially strong forecasts that it permits. This section discusses its relation to different dimensions of resource consumption and environmental impact, and the broader implications of the results in sec.\ \ref{results}. 

First, considering that the exergy needs of the economy are mainly supplied by the so-called \textit{energy production} (which is a meaningless expression in strict thermodynamic terms, but will be used in this section with its ordinary meaning) and the appropriation of biological production (including but not limited to food), the results in sec.\ \ref{results} relate to different resources and impacts to the extent that they are tied to these activities. Second, even for resource losses or environmental impacts with little direct quantitative significance for thermodynamic budgets, there will be thermodynamic costs involved in the recovery or replacement of such resources and the mitigation of or adaptation to such impacts, which could be arbitrarily high depending on the case. Third, even without any physical law making them indispensable, some sets of resources or emissions might be so entangled in the economy that an equation of the same form as Eq.\ \ref{withfloor} should also apply to them, albeit, possibly, with $c_f=0$. In these cases, absolute decoupling would be technically possible, but endogenous innovation would be unable to sustain it (a result from sec.\ \ref{results} that applies both for $c_f > 0$ and for $c_f = 0$). 

Let us focus on energy production. The equations in sec.\ \ref{results} can be used as a basis to tentatively forecast various impacts mediated by such production. For example, let us assume that, during some period, the carbon intensity $\kappa$ of energy supply changes exponentially, $\kappa = \kappa_0 e^{\eta t}$, where $\eta$ could be negative because of technological change or positive because of the decreasing quality of remaining fossil fuels. Then, asymptotically (Eq.\ \ref{assimptotica_abs}), since temperature is roughly proportional to cumulative emissions (Bindoff et al.\, 2013), the expected increase of global temperature attributable to the energy spent in a given (growing) economic activity since $t=0$ would be 
\[ 
\Delta T = \left(\frac{\rho \kappa_0 c_f y_0}{r+ \eta +1}\right) e^{(r+ \eta +1)t} + \left(\frac{\rho \kappa_0 a r^\lambda y_0^{1-\lambda}}{(1-\lambda )r -h +\eta +1}\right) e^{[(1-\lambda )r -h +\eta +1]t}, 
\] 
where $\rho$ is the transient climate response to cumulative CO$_2$ emissions (TCRE), whose likely value is in the range 0.8-2.5 $^o$C per 1000 PgC (Bindoff et al.\, 2013). Section \ref{results} gives the conditions for energy consumption to increase or decrease, and these can also be obtained for emissions. In principle, $\Delta T$ can only increase unless emissions drop to zero (or are compensated by carbon capture), because of its cumulative character (Bindoff et al.\, 2013). By definition, the depletion of nonrenewable resources is also cumulative. 

The significance of the results in sec.\ \ref{results} for the viability of the green growth program largely rely on the viability or not of a sustained and environmentally-friendly growth in energy production (with no implication that this would be a sufficient condition). The complex simulation models used by Hatfield-Dodds et al.\ (2015) in their influential report supporting the viability of green growth in Australia and in the report by the same authors for the whole world (Hatfield-Dodds et al., 2017) do not contradict the results in sec.\ \ref{results} in that they expect growing energy consumption in a growing economy. However, they claim that this is compatible with net greenhouse gas emissions falling to zero, but they rely largely on biosequestration to reach this result, so they switch from increasing warming to increasing occupation of suitable land, obviously a limited resource (which might serve as a temporary relief for Australia but is much scarcer in other countries). Furthermore, climate change is not the only problem posed by fossil fuels (to begin with, their potential cumulative supply is indeed limited, even though the precise amounts and the consequences for noncumulative supply are hotly debated; e.g., Bruckner et al., 2014; Hallock Jr. et al., 2014; Bardi, 2019), and every other energy source has its own constraints. For example, even though renewable sources can supply energy indefinitely, their expansion is limited (Garc{\'i}a-Olivares et al., 2012), by needs such as, again, suitable collection surface (Capell{\'a}n-P{\'e}rez et al., 2017). If indefinite growth is sought, it is unavoidable that renewable energies compete for space with natural ecosystems or with other economic activities. The case of agrofuels is especially problematic, since even their current modest production has already interfered with food supplies (Eide, 2009). Moreover, Lopes Cardozo et al.\ (2016) estimate that massive conversion towards renewables, fusion (should it ever proof workable) or other sources will need a transition phase of several decades during which the supplied energy will be offset by the energy cost of building the industrial capacity to supply it. These problems might be alleviated by emergent solar technologies that, in addition to being little demanding in terms of material resources, are claimed to provide high energy returns on energy investment (Zhou and Carbajales-Dale, 2018), but even in this case the collection surface constraint should ultimately apply. In the longer term, even sources such as hypothetical fusion reactors would cause a problem of excess dissipated heat by-product, should energy consumption keep growing (Berg et al., 2015). 

Even thought, in principle, all energy sources are subject to constraints, the magnitude of such constraints is relevant and often unclear. The literature on this issue is vast, and an exhaustive assessment will not be intended in this paper. However, a point to emphasize is that the prospects for green growth and for growth in general cannot be assessed relying on economic data and theory alone, since, among other inputs, the previous assessment of options for energy supply is essential. It could be replied that completely new energy sources will be developed in the future given the right market signals, so that energy forecasts should rely on economic forecasts rather than the other way round, but this reasoning is flawed for a fundamental reason given in Pueyo (2014, sec.\ 4.1). In that paper a distinction was borrowed from unrelated applications in physics. On the one side, there are problems involving an asymptotically large number of elements, which can and need to be dealt with in statistical terms (in physicists' parlance, these are the cases in which the \textit{thermodynamic limit} is a useful assumption, and are the object of statistical physics; Garrod, 1995). On the other side, there are problems involving a relatively small number of elements, where the best predictions are obtained by considering the behavior of each specific element and their specific interactions. Learning curves, which are central to this paper, synthesize a large number of steps which can only be dealt with statistically, so they belong to the first category\footnote{Therefore the present study is closely related to \textit{ecological econophysics} (Pueyo, 2014), which synthesizes ecological economics with econophysics, i.e., with the application of concepts and tools of statistical physics to economics.}. It is in this category where economic \textit{stylized facts} may be meaningful. In contrast, there are just a handful of energy sources. Therefore, statistical records and economic rules give little information about the prospects for new additions. The possibility of breakthroughs in energy supply is worth discussing, but, being this a \textit{small number} issue, this discussion belongs to engineering, physics and Earth science, with little role for economics. 

While the analysis of energy (exergy) suffices to conclude that long-term economic options cannot be tackled by any economic theory detached from the natural sciences (i.e., by any approach other than ecological economics) and that the prospects for green growth and for growth in general are poor, it should not be forgotten that sustainability is not just a matter of energy, so that many other aspects also merit independent consideration (e.g., Martinez-Alier et al., 1998). Thus far, neither dollars nor joules can bring an extinct species back to life. 

\subsection{Evolving patterns of innovation and limits to assimilable innovation} 
\label{acceleration} 

This section moves beyond the limitations of the model in secs.\ \ref{foundations}-\ref{results} by questioning two implicit assumptions also shared by most of the related literature: (1) That the role of technological development is only positive (also discussed by Pargman and Wallsten, 2017) and (2) that current patterns of innovation (in this case, characterized by Eq.\ \ref{withfloor}) are indefinitely extrapolatable. 

Even if, by applying some given market instruments, eco-innovation becomes a major share of innovation, innovation in other aspects will grow at similar if not greater rates, in some cases just because it might be intertwined with eco-innovation. Policies to facilitate generic innovation make part of the green growth programme, as a support for its more specific tools (UNEP, 2014). Furthermore, for endogenous growth theories, innovation is essential for growth (Barro and Sala-i-Martin, 2004). The basis of exponential growth according to these theories, as well as the result of exponential growth according to most innovation models in sec.\ \ref{foundations}, is accelerating innovation\footnote{In models such as those in sec.\ \ref{foundations}, innovations are generally assumed to result, in part, from experience, whose proxy is cumulative production, and to result in increasing efficiency in terms of some measure, currently monetary costs. If the number of innovations has a linear or a power law relation with cumulative production, or with efficiency while far from its limit if any, exponential production growth will lead to exponential increase in the number of innovations per unit time, i.e., to accelerating innovation, for all the models in sec.\ \ref{foundations}. This would not be true if, e.g., the relation were logarithmic. Some evidence that the former is closer to reality is provided by Magee et al.\ (2016), who found a power law relation of cumulative production and of cost with the number of patents.}. Unfortunately, accelerating innovation results in a number of challenges, which are discussed below. Two implications follow. First, the need to understand and develop institutional solutions to such challenges at a speed commensurate with the rate at which they emerge, which is problematic if innovation accelerates through time. Second, the difficulty to develop such solutions in a system that is based on competition and microeconomic growth imperatives (see Richters and Siemoneit, 2018) rather than solidarity and sufficiency. 

Albeit normally overlooked, the need to adapt the tempo of technological innovation to the tempo of institutional adaptation was already discussed in Meadows et al.\ (1972). These authors gave instances of geographic regions where the Green Revolution had improved people's lives and others in which the social side effects had been clearly detrimental, which they attributed to preexisting institutional differences. In their words, such experiences show that \textit{social side-effects must be anticipated and forestalled} before \textit{the large-scale introduction of a new technology} ({...}). \textit{Such preparation for technological change requires, at the very least, a great deal of time. Every change in the normal way of doing things requires an adjustment time, while the population, consciously or unconsciously, restructures its social system to accommodate the change. While technology can change rapidly, political and social institutions generally change very slowly.} Much more recently, even the literature on long-term implications of artificial intelligence (AI), which, in general, is extremely technophile and pro-growth (Pueyo, 2018), has been warning of the need to slow down the investment in AI development as compared to the investment in means to prevent its potentially irreversible impacts (e.g., Bostrom, 2014), which, however, are thought of as mostly technical means. Arguably, these would be useless in the absence of deep institutional changes (Pueyo, 2018). 

The very challenges that recipes such as green growth pretend to solve were created by technological innovations in combination with economic institutional arrangements, and the adequate institutional changes are indeed lagging behind, whether they are those promoted under the umbrella of \textit{green growth}, \textit{degrowth} or any other. Unfortunately, accelerating innovation is acceleratingly posing new challenges, from an exploding diversity of chemicals with deleterious or yet unknown impacts on health and the environment (UNEP, 2019) to new technologies usable for mass destruction, as was already the case of nuclear energy and appears to be the case of developments not just in AI but also in several other emergent technologies such as nanotechnology or new methods of genetic engineering (see, e.g., Sutherland et al. (2016) for a review of emergent threats, focusing on their significance for biodiversity). If the institutional capabilities to manage such risks does not evolve at a commensurate speed, there will be an accumulation of serious threats for civilization and the rest of the biosphere that will not be solved by increases in eco-efficiency or by new energy sources. 

Subtler social consequences of accelerating innovation are becoming ever more pervasive. Technological acceleration, combined with market competition, are motors of what sociologists call social acceleration (Rosa and Scheuerman, 2009; about the link to innovation, see, e.g., L{\"u}bbe, 2009, p. 169-170), leading some aspects of society to change at accelerating rates and forcing people to adapt. An example is the mounting pressure over workers to increase their adaptability and allied capabilities, collectively known as \textit{employability} (Chertkovskaya et al., 2013). Due to what the economic literature refers to as \textit{bounded rationality}, the human capacity to process change, to adapt and to become adaptable have limits which will be overcome if the pressure grows exponentially. As put by L{\"u}bbe (2009, p. 175), \textit{Processes of growing up, just like processes of growing old, become precarious if the quantity of cultural resources that have consistent validity over the short duration of an average life dissolves with disorienting consequences}. This phenomenon holds some analogy to ecosystem degradation, which, in part, is due to limited capacities to track anthropogenic changes (e.g., Devictor et al., 2012) rather than the inherent implications of, say, a given temperature. 

The pressure over people's lives is not just due to the speed of change in general but also to the fact that much cutting-edge innovation pursues efficiency in the absorption and use of a limited resource as is human attention. This phenomenon was already conceptualized by Simon (1971) as \textit{attention economy} and it is becoming increasingly difficult to escape its influence\footnote{See the udpated information in \url{http://humanetech.com/}}. If models like the one in this paper apply, even approximately, in this case, exponential growth will be linked to a roughly exponential increase in efficiency in the use of this resource (up to saturation), i.e., efficiency in shaping preferences and behavior to the benefit of firms. 

Such considerations become all the most important when taking into account the incipient innovation in the very patterns of innovation. The disruption of traditional innovation curves is underway because of design automation, which has been already going on for some time (Lavagno et al., 2006). In the field of evolvable hardware there are instances of evolutionary algorithms developing highly effective devices whose functioning was elusive for human engineers (Bostrom, 2014, p. 154). The substantial progress that most experts expect for AI after some decades (M\"{u}ller and Bostrom, 2016; Dafoe and Russell, 2016) is likely to result in deep changes in all aspects of society, including the generalization of such \textit{second-order innovation}. When this occurs, the model in this paper and the ensuing results will no longer hold, because that transformation will open the doors to endogenous superexponential innovation. Superexponential innovation is the condition noted in sec.\ \ref{results} to sustain absolute decoupling beyond some point, but only until the physical limit $c_f$ is reached, and if rebound (Schneider, 2008) is avoided. What this entails in terms of ultimate limits to growth will depend on the significance of such enhanced innovation for less clear-cut issues such as those dealt with in sec.\ \ref{energy}. However, if this growth is to be \textit{green} or, more generally, serve the human interest, the problems dealt with above in this same section become even more relevant. Even more important than human capacity to adapt to accelerated changes would be the human limited capacity to take complex decisions, which, in a system based of competition among economic units such as firms, is likely to mean, as noted in Pueyo (2018), an unavoidable, massive automation of decisions, which would become ever more unintelligible and could fully dissociate the directions taken by economies from human interests. 

This section points to some widely ignored but crucial aspects for any economic recipe intended to serve the human interest as green growth is supposed to (as well as for recipes intended to serve the interests of other sentient beings). The first suggestion that emerges is that innovation, however desirable it might be, would need to take place at some pace commensurate with the capacities of people to adapt and of institutions to pre-adapt. Given the strong two-way causal relation generally assumed to exist between innovation and growth, this is a reason to abandon the logic of indefinite growth. The second suggestion is that, given the formidable forces that technology is unleashing and their potential to produce mass destruction or mass dystopia, the most basic institutional change that is needed is a transition from a system based on competition and microeconomic growth imperatives (see Richters and Siemoneit, 2018) to a system that favors solidarity and sufficiency, as previously discussed in Pueyo (2018). Pueyo (2018) also suggests that the broad spectrum of the potentially affected by such impacts creates some hope for collective action in this direction. Such a radical change will also be a form of social acceleration, but transient and oriented by a common purpose, distinct from a state of permanently accelerated and largely meaningless change. 

\section{Conclusions} 
\label{conclusions} 

The aim of this paper was to move beyond conflicting apriorisms and introduce a formal and evidence-based approach to assess the feasibility of absolute decoupling and green growth. Without having reached a final answer, the limited available knowledge appears to give much more reason for skepticism than for confidence, which supports ecological economists' view that mainstream economics' neglect of biophysical dimensions is, at the very least, premature. Considering the tentative results, or just the prevailing uncertainty, the author finds irresponsible the official institutions' messages implicitly presenting the conjecture of green growth as if it were a well-established scientific truth, and, therefore, sidelining other mechanisms to deal with extremely important and urgent global challenges. The paper has focused on the role of technological innovation and learning, leaving changes in GDP composition and their limitations for a separate paper. 

Having noted the little plausibility of technological innovation being quick and effective enough to offset the resource and environmental pressures of growth, it is also suggested that accelerating innovation poses inherent problems requiring institutional changes that can only occur at a limited rate, and that are especially difficult under the existing economic frameworks, which would be perpetuated by green growth policies. The suggestion that follows is for institutions to replace unfounded apriorisms by serious research, and for this research not focusing just on techno-fixes but also on avenues for structural institutional change. Issues of speed, motivating forces of economic activity and growth imperatives need to be seriously considered, as is already attempted in literatures such as degrowth's. 

\section*{Appendix: Value of the learning and innovation exponent $\lambda$} 

A basic assumption in this paper is that the dynamics of monetary cost reductions in current economies have some general features that would be preserved should monetary costs align with thermodynamic costs. Of special interest is the range of values currently taken by the learning exponent $\lambda$ (Eqs.\ \ref{Wright}, \ref{Nordhaus}).

References abound to the \textit{80\% progress curve} (e.g., Dutton and Thomas, 1984), because, in a number of instances, doubling $Q$ leaves $c$ at approximately 80\% the initial value, corresponding to $\lambda \approx 1/3$. To begin with, this was the value found by Wright (1936) for airplane production, in the paper that introduced Eq.\ \ref{Wright}. However, he was quantifying specifically labor costs. Interestingly, he found a much lower value for raw material costs, of $\lambda = 0.0732$, or 95\% curve, meaning more sluggish progress in this aspect (but this result might have been affected by the proximity to a floor cost). Unless otherwise specified, the figures given hereafter refer to total monetary costs. The variability in $\lambda$ is high, but the average of the 108 firm-level instances in figure 2 in Dutton and Thomas (1984) is precisely $\overline{\lambda} \approx 1/3$. However, the 62 industry-level estimates by Nagy et al.\ (2013), which are classified by economic sector, suggest noticeable differences among such sectors, with average $\lambda$ ranging from 0.27 to 0.47.

More important in our case is that all the estimates in these two sources and in many others, including the 331 analyses revised by Weiss et al. (2010), display $\lambda < 1$, as is also generally the case for learning curves in psychology (Newell and Rosenbloom, 1981). The inequality $\lambda < 1$ has sometimes been used as part of the definition of Wright's rule (Anzanello and Fogliatto, 2011). However, not all the estimates in the literature satisfy it. Notably, a large fraction of the 1,938 estimates in Dosi et al.\ (2017) (who use prices as a proxy for costs) display $\lambda > 1$, but the estimates in this sample share some characteristics that may explain this observation. First, these are firm-level values of $\lambda$ (even though the values for different firms are aggregated and an only figure is given for each product). Second, these are firms from an emergent economy in process of \textit{catching up}, India. In view of these two characteristics, much of the change in $c$ is likely to result from the quick incorporation of techniques that were already used elsewhere. Indeed, the large growth rates of emergent economies has been interpreted as a feature of the process of \textit{catching-up}, which would end when they reach the \textit{technological frontier} (Piketty, 2014), and the same effect should be more pronounced for individual firms that have not even reached the technological frontier of their own countries. In the context of the present paper, the main concern is industry-level innovation; furthermore, transient effects are not considered valid to estimate $\lambda$ because they receive separate treatment. At industry level, $\lambda < 1$ is found for the 62 estimates by Nagy et al.\ (2013) and in all the cases reviewed by Weiss et al.\ (2010), which include 132 energy supply technologies, 75 energy demand technologies and 124 other manufacturing processes. Yet another feature of Dosi et al.'s estimates is that they are largely distorted by factors other than cost reductions driven by cumulative production, as noted by the own authors and apparent from the fact that not only a large fraction of estimates correspond to $\lambda > 1$ but there are also a large fraction displaying $\lambda<0$ (decreasing efficiency). Nordhaus (2014) claims, with the help of empirical data, that estimates of $\lambda$ are generally biased upwards and can give spurious results greater than one when the contribution of exogenous innovation is ignored, by using Eq.\ \ref{Wright} instead of Eq.\ \ref{Nordhaus}. For Nordhaus, realistic values of the learning exponent are in the range $0 < \lambda < 0.5$. Mechanisms that may contribute to the seeming ubiquity of Wright's effect with $\lambda < 0.5$, or at least, $\lambda < 1$, are learning-by-doing with diminishing returns, diminishing returns to R\&D effort, firms' R\&D effort increasing with production but decreasing as its returns decrease, and the difficulties to accelerate the implementation of new techniques (including infrastructure renewal and workers' training) as much as to accelerate their development. Given these considerations, in the present paper it is assumed that, as a general rule, $\lambda < 1$.

\section*{Acknowledgments} 

I am grateful, for useful discussions, to the members of Research \& Degrowth, especially Giorgos Kallis, as well as Mart{\'i} Gibert. I thank Centre de Recerca Matem{\`a}tica (CRM) for its hospitality during this work. This research did not receive any specific grant from funding agencies in the public, commercial or not-for-profit sectors.

\section*{References} 

\noindent Anzanello M.J., Fogliatto, F.S., 2011. Learning curve models and applications: Literature review and research directions. Int.\ J.\ Ind.\ Ergonom.\ 41, 573-583.\\ 

\noindent Arora, S., Barak, B., 2009. Computational Complexity: A Modern Approach. Cambridge University Press.\\ 

\noindent Ayres, R.U., van den Bergh, J.C.J.M., Lindenberger, D., Warr, B., 2013. The underestimated contribution of energy to economic growth. Struct.\ Change Econ.\ Dyn.\ 27, 79-88.\\ 

\noindent Bardi, U., 2013. The mineral question: how energy and technology will determine the future of mining. Front.\ Energy Res.\ 1, 9.\\ 

\noindent Bardi, U., 2019. Peak oil, 20 years later: Failed prediction or useful insight? Energy Res.\ Soc.\ Sci.\ 48, 257-261.\\ 

\noindent Barro, R., Sala-i-Martin, X., 2004. Economic Growth, 2nd ed. The MIT Press, Cambridge, MA.\\ 

\noindent Baumg{\"a}rtner, S., 2002. Thermodynamics and the economics of absolute scarcity . Why and how thermodynamics is relevant for ecological, environmental and resource economics, in: Proceedings of the 2nd World Congress of Environmental and Resource Economists, June 24-27, 2002, Monterey, CA.\\

\noindent Berg, M., Hartley, B., Richters, O., 2015. A stock-flow consistent input-output model 
with applications to energy price shocks, interest rates, and heat emissions. New J.\ Phys.\ 17, 015011.\\

\noindent Bindoff, N.L., Stott, P.A., AchutaRao, K.M., Allen, M.R., Gillett, N. et al., 2013: Detection and Attribution of Climate Change: From Global to Regional., in: Stocker, T.F., Qin, D., Plattner, G.-K., Tignor, M., Allen, S.K., et al.\ (eds.) Climate Change 2013: The Physical Science Basis. Contribution of Working Group I to the Fifth Assessment Report of the Intergovernmental Panel on Climate Change. Cambridge University Press.\\ 

\noindent Bithas, K., Kalimeris, P., 2013. Re-estimating the decoupling effect: Is there an actual transition towards a less energy-intensive economy? Energy 51, 78-84.\\ 

\noindent Bostrom, N., 2014. Superintelligence: Paths, Dangers, Strategies. Oxford University Press.\\ 
 
\noindent Bruckner T., Bashmakov, I.A., Mulugetta, Y., Chum, H., de la Vega Navarro, A. et al., 2014. Energy Systems, in: Edenhofer, O., Pichs-Madruga, R., Sokona, Y., Farahani, E., Kadner, S. et al.\ (Eds.), Climate Change 2014: Mitigation of Climate Change. Contribution of Working Group III to the Fifth Assessment Report of the Intergovernmental Panel on Climate Change. Cambridge University Press.\\ 

\noindent Capell{\'a}n-P{\'e}rez, I., de Castro, C., Arto, I., 2017. Assessing vulnerabilities and limits in the transition to renewable energies: Land requirements under 100\% solar energy scenarios. Renew.\ Sustain.\ Energy Rev.\ 77, 760-782.\\ 

\noindent Lopes Cardozo, N.J., Lange, A.G.G., Kramer, G.J., 2016. Fusion: Expensive and taking forever? J.\ Fusion Energy 35, 94-101.\\ 

\noindent Cattaneo, C., Dittmer, K., D'Alessandro, S., Demaria, F., Domene, E., F{\'i}guls, M., Galletto, V., Garc{\'i}a, M., Marull, J., P{\'e}rez, A., Porcel, S., Sorman, A., 2016. A Study on Job Creation in a Post‐Growth Economy. The Greens - EFA, Green European Foundation.\\ 

\noindent Chertkovskaya, E., Watt, P., Tramer, S., Spoelstra, S., 2013. Giving notice to employability. Ephemera 13, 701-716.\\ 

\noindent Conrad, J., 1995. Greenfreeze: environmental success by accident and strategic action, in: J{\"a}nicke, M., Weidner, H.\ (Eds.), Successful Environmental Policy: A Critical Evaluation of 24 Cases. Sigma, Berlin, pp.\ 364-378.\\ 

\noindent Csereklyei, Z., Stern, D.I., 2015. Global energy use: Decoupling or convergence? Energy Econ.\ 51, 633-641.\\ 

\noindent Dafoe, A., Russell, S., 2016. Yes, we are worried about the existential risk of artificial intelligence. MIT Technol.\ Rev., 2 November. \url{https://www.technologyreview.com/s/602776/yes-we-are-worried-about-the-existential-risk-of-artificial-intelligence/}\\ 

\noindent Dale, G., Mathai, M.V., Oliveira, J.A.P.\ (Eds.), 2016. Green Growth : Ideology, Political Economy and the Alternatives. Zed Books, London.\\ 

\noindent D'Alisa, G., Demaria, F., Kallis, G.\ (Eds.), 2014. Degrowth: a Vocabulary for a New Era. 
Routledge, London.\\ 

\noindent Daly, H.E.\ (Ed.), 1973. Toward a Steady-State Economy. W.H. Freeman and Co., San Francisco.\\

\noindent Devictor, V., Van Swaay, C., Brereton, T., Brotons, L., Chamberlain, D., et al., 2012. Differences in the climatic debts of birds and butterflies at a continental scale. Nat.\ Clim.\ Change 2, 121-124.\\ 

\noindent Dincer, I., Rosen, M.A., 2013. Exergy: Energy, Environment and Sustainable Development, 2nd ed. Elsevier, Oxford.\\ 

\noindent Donner, Y., Hardy, J.L., 2015. Piecewise power laws in individual learning curves. Psychon.\ Bull.\ Rev.\ 22, 1308-1319.\\ 

\noindent Dosi, G., Grazzi, M., Mathew, N., 2017. The cost-quantity relations and the diverse patterns of “learning by doing”: Evidence from India. Res.\ Policy 46, 1873-1886.\\ 

\noindent Dutton, J.M. and Thomas, A., 1984. Treating progress functions as a managerial opportunity . Acad.\ Manag.\ Rev.\ 9, 235-247.\\ 

\noindent Eide, A., 2009. The Right to Food and the Impact of Liquid Biofuels (Agrofuels). FAO, Rome.\\ 

\noindent Garc{\'i}a-Olivares, A., Ballabrera-Poy, J., Garc{\'i}a-Ladona, E., Turiel, A. 2012 A global renewable mix with proven technologies and common materials. Energy Policy 41, 561-574.\\ 

\noindent Garrod, C., 1995. Statistical Mechanics and Thermodynamics. Oxford University Press.\\ 

\noindent Georgescu-Roegen, N., 1971. The Entropy Law and the Economic Process. Harvard University Press, Cambridge, MA.\\ 

\noindent Haas, W., Krausmann, F., Wiedenhofer, D., Heinz, M., 2015. How circular is the global economy? An assessment of material flows, waste production, and recycling in the european union and the world in 2005. J.\ Ind.\ Ecol.\ 19, 765-777.\\ 

\noindent Hallock Jr., J.L., Wu, W., Hall, C.A.S., Jefferson, M., 2014. Forecasting the limits to the availability and diversity of global conventional oil supply: Validation. Energy 64, 130-153.\\ 

\noindent Hardt, L., Owen, A., Brockway, P., Heun, M.K., Barrett, J., Taylor, P.G., Foxon, T.J., 2018. Untangling the drivers of energy reduction in the UK productive sectors: Efficiency or offshoring? Appl.\ Energy 223, 124-133.\\ 

\noindent Hatfield-Dodds, S., Schandl, H., Adams, P.D., Baynes, T.M., Brinsmead, T.S., Bryan, B.A., et al., 2015. Australia is 'free to choose' economic growth and falling environmental pressures. Nature 527, 49-53.\\ 

\noindent Hatfield-Dodds, S., Schandl, H., Newth, D., Obersteiner, M., Cai, Y., Baynes, T., et al., 2017. Assessing global resource use and greenhouse emissions to 2050, with ambitious resource efficiency and climate mitigation policies. J.\ Clean.\ Prod.\ 144, 403e414.\\ 

\noindent Hickel, J., 2019. Degrowth: a theory of radical abundance. Real World Econ.\ Rev.\ 87, 54-68.

\noindent IEA (International Energy Agency), 2007. Tracking Industrial Energy Efficiency and CO$_2$ Emissions. OECD/IEA, Paris.\\ 

\noindent Kallis, G., 2017. Radical dematerialization and degrowth. Phil.\ Trans.\ R. \
Soc.\ A 375, 20160383.\\

\noindent Kallis, G., 2018. Degrowth. Agenda Publishing, Newcastle upon Tyne, UK.\\ 

\noindent Kallis, G., Kostakis, V., Lange, S., Muraca, B., Paulson, S., Schmelzer, M., 2018. Research On Degrowth. Annu.\ Rev.\ Environ.\ Resour.\ 43, 291-316\\

\noindent Kallis, G., Sager, J., 2017. Oil and the economy: A systematic review of the literature for ecological economists. Ecol.\ Econ.\ 131, 561-571.\\ 

\noindent K{\"o}ler, J., Grubb, M., Popp, D., Edenhofer, O., 2006. The transition to endogenous technical change in climate-economy models: A technical overview to the Innovation Modeling Comparison Project. Energy J.\ 27, 17-55.\\ 

\noindent K{\"u}mmel, R., Lindenberger, D., 2014. How energy conversion drives economic growth far from the equilibrium of neoclassical economics. New J.\ Phys.\ 16, 125008 

\noindent Lavagno, L., Martin, G., Scheffer, L., 2006. Electronic Design Automation for Integrated 
Circuits Handbook. CRC Press.\\ 

\noindent L{\"u}bbe, H., 2009. The contraction of the present, in: Rosa, H., Scheuerman, W.E.\ (Eds.), High-speed Society. Pennsylvania State University Press, pp. 159-178  (Translation from: L{\"u}bbe, H., 1998. Gegenwartsschrumpfung, in: Backhaus, K., Bonus, H.\ (Eds.), Die Beschleunigungsfalle oder der Triumph der Schildkr{\"o}te, Sch{\"a}fer-P{\"o}schel, Stuttgart).\\ 

\noindent Magee, C.L., Devezas, C., 2017. A simple extension of dematerialization theory: Incorporation of technical progress and the rebound effect. Technol.\ Forecast.\ Soc.\ Chang.\ 117, 196-205.\\ 

\noindent Magee, C.L., Devezas, C., 2018. Specifying technology and rebound in the IPAT identity. Procedia Manuf.\ 21, 476-485.\\ 

\noindent Magee, C.L., Basnet, S., Funk, J.L., Benson, C.L., 2016. Quantitative empirical trends in technical performance . Technol.\ Forecast.\ Soc.\ Chang.\  104, 237-246.\\ 

\noindent Martinez-Alier, J., 2013. Ecological Economics, in: International Encyclopedia of the Social and Behavioral Sciences, Elsevier, Amsterdam, p. 91008.\\

\noindent Martinez-Alier, J., Munda, G., O'Neill, J., 1998. Weak comparability of values as a foundation for ecological economics. Ecol.\ Econ.\ 26, 277-286.\\ 

\noindent Meadows, D.H., Meadows, D.L., Randers, J., Behrens, W.W., 1972. The Limits to Growth. Universe Books, New York.\\ 

\noindent M\"{u}ller, V.C. and Bostrom, N. 2016. Future progress in artificial intelligence: A survey of expert opinion, in: M\"{u}ller, V.C.\ (Ed.), Fundamental Issues of Artificial Intelligence, Springer, Berlin, pp. 553-571.\\ 

\noindent Nagy, B., Farmer, J.D., Bui, Q.M., Trancik, J.E., 2013. Statistical basis for predicting technological progress. PLoS One 8, e52669.\\ 

\noindent Newell, A., Rosenbloom, P. S., 1981. Mechanisms of skill acquisition and the law of practice., in: Anderson, J.R.\ (Ed.), Cognitive skills and their acquisition, vol. 6. Erlbaum, Hillsdale, NJ, pp. 1-55.\\ 

\noindent Nordhaus, W.D., 2014. The perils of the learning model for modeling endogenous 
technological change. Energy J.\ 35, 1-13.\\ 

\noindent OECD, 2011a. Towards Green Growth . A Summary for Policy Makers. OECD, Paris.\\ 

\noindent OECD, 2011b. Tools for Delivering on Green Growth. OECD, Paris.\\ 

\noindent Pargman, D., Wallsten, B., 2017. Resource scarcity and socially just internet access over time and space, in LIMITS 2017 - Proceedings of the 2017 Workshop on Computing Within Limits. Association for Computing Machinery, Inc., New York, pp. 29-36. \url{https://dl.acm.org/citation.cfm?doid=3080556.3084083}\\ 

\noindent Piketty, T. 2014. Capital in the Twenty-First Century. Harvard University Press, Cambridge, MA.\\ 

\noindent Plank, B., Eisenmenger, N., Schaffartzik, A., Wiedenhofer, D., 2018. International trade drives global resource use: A structural decomposition analysis of raw material consumption from 1990-2010. Environ.\ Sci.\ Technol.\ 52, 4190-4198.\\ 

\noindent Pueyo, S., 2003. Irreversibility and Criticality in the Biosphere. University of Barcelona. \url{http://hdl.handle.net/2445/35290}\\ 

\noindent Pueyo, S., 2014. Ecological econophysics for degrowth. Sustainability 6, 3431-3483.\\ 

\noindent Pueyo, S., 2018. Growth, degrowth, and the challenge of artificial superintelligence. J.\ Clean.\ Prod.\ 197, 1731-1736.\\ 

\noindent Ram{\'i}rez, C.A., Worrell, E., 2006. Feeding fossil fuels to the soil. An analysis of energy embedded and technological learning in the fertilizer industry. Conserv. Recycl. 46, 75-93.\\ 

\noindent Richters, O., Siemoneit, A., 2018. The contested concept of growth imperatives: Technology and the fear of stagnation. Oldenburg Discussion Papers in Economics, V-414-18.\\

\noindent Sahal, D., 1979. A theory of progress functions. AIIE Trans.\ 11, 23-29.\\ 

\noindent Rosa, H., Scheuerman, W.E., 2009. High-speed Society. Pennsylvania State University Press.\\ 

\noindent Schandl, H., Fischer-Kowalski, M., West, J., Giljum, S., Dittrich, M., Eisenmenger, N., et al., 2018. Global material flows and resource productivity . Forty years of evidence. J.\ Ind.\ Ecol.\ 22, 827-838.\\ 

\noindent Schneider, F., 2008. Macroscopic rebound effects as argument for economic degrowth. Proceedings of the First International Conference on Economic De-growth for Ecological Sustainability and Social Equity, Paris, April 18-19th 2008. \url{http://events.it-sudparis.eu/degrowthconference/en/themes/1First\%20panels/Backgrounds/Schneider\%20F\%20Degrowth\%20Paris\%20april\%202008\%20paper.pdf}\\ 

\noindent Simon, H.E. 1971. Designing organizations for an information-rich world, in: Greenberger, M.\ (Ed.), Computers, Communications, and the Public Interest. The Johns Hopkins Press. Baltimore, MD, pp. 37-52.\\

\noindent Smulders, S., Toman, M., Withagen, C., 2014. Growth theory and 'green growth'. Oxford Rev.\ Econ.\ Policy 30, 423-446.\\ 

\noindent Soddy, F., 1922. Cartesian Economics. The Bearing of Physical Science upon State Stewardship. Hendersons, London.\\ 

\noindent Solow, R.M., 1973. Is the End of the World at hand? Challenge, 16(1), 39-50.\\ 

\noindent Steinberger, J.K., Krausmann, F., Eisenmenger, N., 2010. Global patterns of materials use: A socioeconomic and geophysical analysis. Ecol.\ Econ.\ 69, 1148-1158.\\ 

\noindent Steinberger, J.K., Krausmann, F., Getzner, M., Schandl, H. and West, J., 2013. Development and dematerialization: An international study. PLoS ONE 8, e70385.\\ 

\noindent Sutherland, W.J., Broad, S., Caine, J., Clout, M., Dicks, L.V., et al., 2016. A horizon scan of global conservation issues for 2016. Trends Ecol.\ Evol.\ 31, 44-53.\\ 

\noindent Tollefson, J., 2009. Cutting out the chemicals. Nature 457, 518-519.\\ 

\noindent UNEP, 2014. Decoupling 2: technologies, opportunities and policy options. United Nations Environment Programme, Nairobi.\\ 

\noindent UNEP, 2019. Global Chemicals Outlook II. United Nations Environment Programme, Nairobi.\\ 

\noindent Ward, J.D., Sutton, P.C., Werner, A.D., Costanza, R., Mohr, S.H., Simmons, C.T., 2016. Is decoupling GDP growth from environmental impact possible? PLoS ONE 11, e0164733.\\ 

\noindent Weiss, M., Junginger, M., Patel, M.K. and Blok, K., 2010. A review of experience curve analyses for energy demand technologies. Technol.\ Forecast.\ Soc.\ Chang.\ 77, 411-428.\\ 

\noindent Wright, T.P., 1936. Factors affecting the cost of airplanes. J.\ Aeronaut.\ Sci.\ 3, 122-128.\\ 

\noindent Zhou, Z., Carbajales-Dale, M., 2018. Assessing the photovoltaic technology landscape: efficiency and energy return on investment (EROI). Energy Env.\ Sci.\ 11, 603-608. 
 
\end{document}